
\magnification=\magstephalf
\baselineskip=16pt
\vsize=9.0 true in


\def\cf{{\it cf. }}
\def\etal{{\it et al. }}

\def\aproxgt{\mathrel{%
	\rlap{\raise 0.511ex \hbox{$>$}}{\lower 0.511ex \hbox{$\sim$}}}}
\def\aproxlt{\mathrel{%
	\rlap{\raise 0.511ex \hbox{$<$}}{\lower 0.511ex \hbox{$\sim$}}}}

\centerline{\bf A Cautionary Note on Gamma Ray Burst Nearest Neighbor
Statistics}
\medskip
\centerline{\sl Michael A. Nowak}
\centerline{\sl CITA, 60 St. George St., Toronto, Ontario ~M5S
1A7}
\centerline{\sl Received 1993 August 6; accepted 1993 ---}

\medskip

\centerline{ABSTRACT}
\medskip
\baselineskip=12pt
In this letter we explore the suggestion of Quashnock and Lamb (1993)
that nearest neighbor correlations among gamma ray burst positions
indicate the possibility of burst repetitions within various burst
sub-classes.  With the aid of Monte Carlo calculations we compare the
observed nearest neighbor distributions with those expected from an
isotropic source population weighted by the published BATSE exposure
map.  The significance of the results are assessed via the
Kolmogorov-Smirnov (K-S) test, as well as by a comparison to Monte
Carlo simulations. The K-S results are in basic agreement with those of
Quashnock and Lamb.  However, as Narayan and Piran (1993) point out,
and the Monte Carlo calculations confirm, the K-S test overestimates
the significance of the observed distributions. We compare the
sensitivity of these results to both the definitions of the assumed
burst sub-classes and the burst positional errors.  Of the
two, the positional errors are more significant and indicate that the
results of Quashnock and Lamb may be due to systematic errors, rather
than any intrinsic correlation among the burst positions.  Monte
Carlo simulations also show that with the current systematic
errors, the nearest neighbor statistic is not very sensitive to
moderate repetition rates. Until the BATSE statistical and systematic
errors are fully understood, the burst nearest neighbor correlations
cannot be claimed to be significant evidence for burst repetitions.
\smallskip
\centerline
{{\sl Subject Headings:} ~  gamma rays: bursts ~---~ methods:
statistical}

\baselineskip=16pt

\bigskip \centerline{1. Introduction}\smallskip

Currently the three most favored scenarios for the
location of gamma ray bursts are that they originate in the Oort
cloud, an extended galactic halo, or at cosmological distances (for
recent reviews, see Blaes 1993, Hartmann 1994).  If bursts are found
to repeat from the same location in the sky, this would have profound
implications for which of these three models is correct.  For
example, it is difficult to imagine an Oort cloud model that allows
repetitions, whereas a galactic halo model
could allow for several events from a single neutron star (see the
aforementioned reviews for references to specific models). In the
cosmological model, burst repetitions can be produced by the effects
of gravitational lensing, although these should be relatively rare
events (Paczy\'nski 1986; Mao 1992, 1993; and others).

One way to search for repetitions and other anisotropies
is to perform a nearest neighbor analysis and search for
excesses at small angles on the order of the BATSE statistical and
systematic errors.  That is, one can ask the question: \lq\lq For a
randomly chosen burst, what is the probability that its nearest
neighbor lies within an angle $\theta$, and how does that probability
compare with what is expected from an isotropic, non-repeating
distribution?"  Quashnock and Lamb (1993) (QL93 hereafter) have
addressed this question and claim to find a burst
excess at small angular scales. Here we investigate the
rigorousness of this result and try to determine whether the observed
deviations from the expected distributions can be attributed to an
intrinsic property of the sources.

\medskip
\centerline{2. Methods}
\smallskip

As QL93 point out, if $(N + 1)$ gamma ray bursts are drawn from an
isotropic distribution, then
$$P(y) = 1 - \exp (- N y/2 ) \eqno(1)$$
is the probability that any given burst has at least one neighbor
that lies within an angle $\theta$ such that $1 - \cos \theta \le y$.
This formula does not take into account the detector response,
therefore we use a Monte Carlo approach instead.
Random samples of $(N+1)$ bursts are drawn from an isotropic
distribution weighted by the detector response (calculated
from a curve fit to the azimuthally averaged BATSE exposure map;
Fishman \etal 1992).  Several hundred of these random samples are
averaged together to yield the theoretical distributions.

The theoretical distributions are compared to distributions
calculated from the publicly available burst data set (Fishman
\etal 1992) which currently contains positions for 260 bursts.
Following QL93, the bursts are divided into two categories, \lq\lq
Type I" and \lq\lq Type II", based upon a measure of their
variability (\cf Lamb \etal 1993; Lamb and Graziani 1993a,b).
Defining $V = (\overline C^{64} )_{max}/ (\overline C^{1024} )_{max}$, where
$(\overline C^{64} )_{max}$ and $(\overline C^{1024} )_{max}$ are respectively
the maximum count rate in 64 ms and 1024 ms, Type I bursts
have $\log V \le -0.8$, and Type II bursts have $\log V > -0.8$.
Type I is found to contain 160 bursts, while Type II is found to
contain 44 bursts.  (V is not a well defined quantity for all 260
bursts.) This is not the only basis for subdividing the bursts (\cf
Kouveliotou \etal 1993), and variations are explored in this work.
Specifically, \lq\lq Type Ia" will be defined to have $\log V \le
-0.7$ and  \lq\lq Type IIa" will be defined to have $\log V > -0.7$.
Type Ia contains 167 bursts and Type IIa contains 37 bursts. \lq\lq
Type Ir" and \lq\lq Type IIr" are based upon a completely random
subdivision of those bursts with a defined $V$. 160 bursts are
assigned to Type Ir and 44 bursts are assigned to Type IIr.  These
other subdivisions are defined in order to explore the sensitivity of
the results to the classifications. All three varieties of Type I
bursts are further subdivided into three roughly equal groups -- Faint,
Medium, and Bright -- ordered by their maximum count rate in 1024 ms.

QL93 compare the data distributions to the theoretical
distributions with the aid of the Kolmogorov-Smirnov (K-S) test (Press
\etal 1986), which is a measure of the maximum deviation of the
measured distribution from the theoretical distribution.  This
deviation is assigned a  \lq\lq $Q$" value, which is a measure of the
likelihood that the two  distributions are the same.  However, as
Narayan and Piran (1993) recently have shown, the K-S test actually
overestimates the true significance of an overabundance of nearest
neighbors.  This is because the K-S test assumes uncorrelated data
points, whereas nearest neighbors tend to become correlated as their
separations decrease.  A more proper statistic can be calculated
from Monte Carlo simulations of the maximum deviations between the
measured distributions and the theoretical distributions. (In this
paper the significance of the deviations is estimated from 10,000
Monte Carlo simulations of each theoretical distribution.)  The $Q$
from the K-S test ($Q_{KS}$) are {\it always} smaller than the $Q$
from the Monte Carlo simulations ($Q_{M}$), and to a good approximation
$Q_{M} \sim {Q_{KS}}^{0.7}$.

The above statistics do not account for
experimental errors or uncertainty.  First, there is the worry that
systematic errors may actually {\it produce} a false signal. Second,
whereas one expects averaging the burst positions over the error boxes
to degrade the signal's significance, the measured significance
should be near the \lq\lq mean" significance.  A mean
significance near unity and a measured significance
several standard deviations away from this mean could indicate
several possibilities, such as systematic errors producing a false
signal, or a very unusual chance occurrence of a false signal. At the
very least such an occurence indicates that the errors need to be
studied in detail. The public data set contains one sigma error
bars that give the radius, in degrees, of a circle with the same area
as the calculated BATSE one sigma error ellipse (Fishman \etal 1992).
In this work the error box is taken to be gaussianly distributed in
right ascension and declination, with the one sigma square having the
same area as the {\it larger} of the BATSE error ellipse or a circle of
radius $4^\circ$ (the BATSE systematic error).  Each burst data set is
then varied with these errors in mind, and the mean and standard
deviation of $\log Q_{M}$ are calculated for each burst sub-class.

\medskip
\centerline{3. Results}
\smallskip
The results of the calculations are presented in Table 1.
With the proper Monte Carlo estimation of the significance, only the
All Type I \& II and All Bursts sub-classes deviate
significantly from the expected distribution ($Q_{M} = 2.4 \times
10^{-3}$, $8.5 \times 10^{-3}$, respectively).  Bright Type I and Ia
and the Faint \& Bright Type Ir subclasses are marginally significant
($Q_{M} \sim 1 - 3 \times 10^{-2}$). The $Q_{KS}$ values, however,
are in rough agreement with those of QL93 (who take slightly
different subdivisions of the Type I class). In
agreement with QL93, the maximum deviations between the measured and
theoretical distributions tend to be burst excesses that occur at an
angular scale of $4^\circ - 5^\circ$, the BATSE systematic error. All
varieties of Bright Type I bursts, however, have deficits at angles
near $20^\circ$. Plots of $P(y)$ for two of the measured and
theoretical distributions are presented in Figures 1a,b.

It is questionable if the Type I -- Type II division plays any role in
determining the deviations from the expected distributions.  None of
the individual Type I or Type II sub-classes shows significant
deviations on its own.  In addition, the third most significant
measurement (Bright Type Ia) was based on a different definition than
that of QL93, and the fourth most significant measurement (Faint \&
Bright Type Ir) was based on a random subdivision. There is the
worry that given enough \lq\lq reasonable" subdivisions,
some fraction of them must show \lq\lq significant" deviations.

The results of the Monte Carlo simulations show that all of the burst
sub-classes, with the exceptions of Faint Type Ia and the Faint
Type Ir, have measured $\log Q_M$ that are less than the mean
($\overline{\log Q_M}$).  Furthermore, most of the means are within
$1.2$ standard deviations ($\sigma_{\log Q_M}$) of $\log Q_M = 0$, with
only the Bright Type I sub-classes falling more than 2 standard
deviations from $0$.   Histograms for the distribution of $\log Q_M$
for two sub-classes are presented in Figures 1a,b.  In general, all of
the sub-classes have $\log Q_M$ histograms that are reasonably
consistent with no deviations from an isotropic, non-repeating
distribution.

This result is somewhat surprising in light of the apparent
significance of the $Q_M$ for the All Type I \& II  and
All Bursts sub-classes.  Figure 1b shows that the measured $\log Q_M$
for the former sub-class falls on the tail of the histogram ($4.6
{}~\sigma_{\log Q_M}$ from the mean).  In addition, the measured $\log
Q_M$ for All Bursts falls $2.6 ~\sigma_{\log Q_M}$ from the mean, and
the measured $\log Q_M$ for the various Faint \& Bright Type I
sub-classes fall $2 - 3 ~\sigma_{\log Q_M}$ from the mean.  This is an
unlikely situation if the errors are solely statistical.  Without
detailed knowledge of the shapes of the BATSE error boxes, however, any
preferential clustering or declustering cannot be modeled. Of the
most significant measurements, only the various Bright Type I
bursts have measured $\log Q_M$ that fall within $2 ~\sigma_{\log
Q_M}$ of the mean. However, these measurements show burst deficits at
$\sim 20^\circ$ rather than burst excesses at $\sim 4^\circ$, and they
are only marginally significant.

As an illustration of the above points, Figure 2 presents the results
of a Monte Carlo simulation where it was assumed that each burst has a
$20\%$ probability of repeating once within the data set.  Bursts
were generated and checked for repetition until a total of 204 bursts
was reached.  This process was repeated for 100 data sets. A $4^\circ$
systematic error (distributed as described above) was applied to each
set and the logarithm of the significance value was calculated ($\log
Q_R$ in Figure 2).  This was repeated 100 times.  Each of these
simulations in turn had a further $4^\circ$ systematic error applied
and the mean logarithm of the significance value was calculated ($\log
Q_{RR}$ in Figure 2).  This was done to simulate the posterior
averaging over error boxes that we performed on the measured
distributions.  Figure 2 plots $\log Q_R$ vs. $\log Q_{RR}$ for these
10,000 simulations.  There are two things to note here.  First, even
with a $20 \%$ chance of repetition, the average $\log Q_R$ is greater
than $-2$, indicating that the nearest neighbor test is not very
sensitive in light of the systematic errors.  Second, the measured
$\overline {\log Q_M}$ (corresponding to $\log Q_{RR}$) are unusually
close to $0$.  Only $2$ out of $10,000$ simulations have both $\log
Q_R < \log Q_M$ and $\log Q_{RR} > \overline {\log Q_M}$, where the
measured values are for the $204$ Type I \& II Bursts with {\it only}
the $4^\circ$ systematic error applied.  Again, the measurements that
are most consistent with the  distribution are the results for the
various Bright Type I sub-classes, which show burst deficits not
excesses.

\medskip
\centerline{4. Conclusions}
\smallskip

We have reexamined the analysis of Quashnock and Lamb (1993) in order
to determine whether or not the measured gamma ray burst nearest
neighbor distributions differ significantly from the theoretical
distributions.  Following QL93, the bursts were subdivided into
several sub-classes which deviated from the theoretical distributions
with marginal significance.  It is unclear to what extent these
significances were effected by the definition of the sub-classes.

Varying the data by the larger of the
systematic or published errors showed that the logarithm of
the measured significance values, $\log Q_M$, were typically lower than
their means.  The smallest (\lq\lq most significant")
measured $\log Q_M$ were seen to be $\sim 2 - 4.6$ standard deviations
away from this mean.  This is an unlikely situation for a purely
statistical error, however, it might be possible to explain with a
systematic error.  The most \lq\lq robust" results (Bright
Type I Bursts) showed burst deficits at $20^\circ$ rather than burst
excesses at $4^\circ$.

Monte Carlo simulations that include both single and double
application of the systematic errors show the extent to which the
experimental results are unusual. They also show that the
nearest neighbor statistic is not very sensitive when $20 \%$ or
fewer of the bursts repeat. We end by reiterating the warning that
without an understanding of the burst positional errors, the burst
nearest neighbor correlations cannot be claimed to be significant
evidence for burst repetitions or clustering.

\bigskip
\centerline{ACKNOWLEDGEMENTS}
\smallskip
I would like to acknowledge useful discussions with Lars Bildsten, Omer
Blaes, Wlodek Kluzniak, Prasenjit Saha, Chris Thompson, Scott Tremaine,
and John Wang.

\bigskip
\centerline{REFERENCES}
\medskip

\noindent\hangindent=20pt\hangafter=1{
Blaes, O.~M. ~1993, ApJS, to be published.}

\noindent\hangindent=20pt\hangafter=1{
Fishman, G.~J., et al. ~1992, BATSE Burst Catalog, GROSSC.}

\noindent\hangindent=20pt\hangafter=1{
Hartmann, D.~H. ~1994, to be published in \lq\lq High Energy
Astrophysics", ed. J. Matthews (World Scientific).}

\noindent\hangindent=20pt\hangafter=1{
Kouveliotou, C., et al. ~1993, ApJ, submitted.}

\noindent\hangindent=20pt\hangafter=1{
Lamb, D.~Q., and Graziani, C. ~1993a, ApJ, in press.}

\noindent\hangindent=20pt\hangafter=1{
Lamb, D.~Q., and Graziani, C. ~1993b, ApJ, in press.}

\noindent\hangindent=20pt\hangafter=1{
Lamb, D.~Q., Graziani, C., and Smith, I.~A. ~1993, ApJ, 413, L11.}

\noindent\hangindent=20pt\hangafter=1{
Mao, S. ~1992, ApJ, 389, L41.}

\noindent\hangindent=20pt\hangafter=1{
Mao, S. ~1993, ApJ, 402, 382.}

\noindent\hangindent=20pt\hangafter=1{
Narayan, R., and Piran, T. ~1993, MNRAS, submitted.}

\noindent\hangindent=20pt\hangafter=1{
Paczy\'nski, B. ~1986, ApJL, 308, L43.}

\noindent\hangindent=20pt\hangafter=1{
Press, W., et al. ~1986, Numerical Recipes (Cambridge: Cambridge
University Press), p. 472.}

\noindent\hangindent=20pt\hangafter=1{
Quashnock, J.~M., and Lamb, D.~Q. ~1993, MNRAS, submitted.}

\vfill\eject

\centerline{\bf Table 1:  Burst Correlation Significance Values}
\bigskip

\def\smalline{height8pt&\omit&\omit&&\omit&&\omit&&\omit&&\omit&\cr}
\def\hrline{\multispan{12}\hrulefill\cr}

\centerline{\vbox{
{\offinterlineskip
\halign{\vrule # & \strut \ \ \sl # \    \hfil
& \hfil # \ & \vrule # & \ \ \ # \  \hfil & \vrule # &
\strut \ \ \  # \ \  \hfil & \vrule # & \hfil \  \ # \
& \vrule # & \hfil \ \  \ # \ \
& \vrule # \cr
\hrline
\smalline
&{\rm Data Set} & (\# bursts) & & $Q_{KS}$ & & $Q_{M}$ & &
 $\overline{\log Q_{M}}$ & & $\sigma_{\log Q_{M}}$  & \cr
\hrline
\smalline
&Faint \lq\lq Type I" & (53) & & $3.6 \times 10^{-2}$
   & & $0.12$ & & $-0.54$ & & $0.47$ & \cr
&Middle \lq\lq Type I" & (54) & & $6.2 \times 10^{-2}$
   & & $0.16$ & & $-0.46$ & & $0.40$ & \cr
&Bright \lq\lq Type I" & (53) & & $4.6 \times 10^{-3}$
   & & $2.5 \times 10^{-2}$ & & $-1.25$ & & $0.57$ & \cr
&\lq\lq Type II" & (44) & & $0.21$ & & $0.40$ & & $-0.42$ & & $0.38$ &
   \cr
&Faint \& Bright \lq\lq Type I" & (106) & & $1.4 \times 10^{-2}$
   & & $5.8 \times 10^{-2}$ & & $-0.46$ & & $0.38$  & \cr
\smalline
\hrline
\smalline
&Faint \lq\lq Type Ia" & (56) & & $0.20$ & & $0.39$ & & $-0.49$ & &
   $0.43$ & \cr
&Middle \lq\lq Type Ia" & (55) & & $0.11$ & & $0.26$ & & $-0.58$
   & & $0.47$ & \cr
&Bright \lq\lq Type Ia" & (56) & & $1.6 \times 10^{-3}$
   & & $1.2 \times 10^{-2}$ & & $-1.32$ & & $0.58$  & \cr
&\lq\lq Type IIa" & (37) & & $0.15$ & & $0.31$ & & $-0.40$
   & & $0.38$ & \cr
&Faint \& Bright \lq\lq Type Ia" & (112) & & $1.4 \times 10^{-2}$
   & & $5.5 \times 10^{-2}$ & & $-0.45$ & & $0.38$  & \cr
\smalline
\hrline
\smalline
&Faint \lq\lq Type Ir" & (53) & & $0.26$ & & $0.47$ & & $-0.45$ & &
   $0.40$ & \cr
&Middle \lq\lq Type Ir" & (54) & & $0.11$ & & $0.24$ & & $-0.50$ & &
   $0.41$ & \cr
&Bright \lq\lq Type Ir" & (53) & & $1.1 \times 10^{-2}$
   & & $4.8 \times 10^{-2}$ & & $-1.23$ & & $0.62$  & \cr
&\lq\lq Type IIr" & (44) & & $0.20$  & & $0.39$ & & $-0.40$ & &
   $0.35$ & \cr
&Faint \& Bright \lq\lq Type Ir" & (106) & & $3.5 \times 10^{-3}$
   & & $2.3 \times 10^{-2}$ & & $-0.45$ & & $0.40$  & \cr
\smalline
\hrline
\smalline
&All \lq\lq Type I \& II" & (204)  & & $9.7 \times 10^{-5}$ & & $2.4
   \times 10^{-3}$ & & $-0.51$ & & $0.46$  &\cr
\smalline
\hrline
\smalline
&All Bursts & (260) & & $9.7 \times 10^{-4}$ & & $8.5 \times 10^{-3}$ &
   & $-0.61$ & & $0.50$ &\cr
\smalline
\hrline}}}}
\bigskip

Table 1:  The significance values that each burst sub-class with the
given \# of bursts is drawn from an isotropic distribution weighted by
the BATSE exposure map.  $Q_{KS}$ are estimated from the
Kolmogorov-Smirnov test, and the $Q_M$ are estimated by
comparison to 10,000 Monte Carlo simulations.  The mean of the
logarithm of this value ($\overline{\log Q_M}$) and the standard
deviation of the logarithm ($\sigma_{\log Q_M}$), are based upon 1000
Monte Carlo calculations that include the burst positional
uncertainty.

\vfill\eject
\centerline{Figure Captions}
\bigskip

Figure 1a -- {\it Top}: Measured cumulative distribution, $P(y)$, for
the 53 brightest Type I bursts, and Monte Carlo calculation for the
theoretical distribution, $P(y)$, for 53 bursts drawn from an
isotropic distribution weighted by the BATSE exposure map.

{\it Bottom}:  Histogram of $\log Q_M$, the logarithm of the
significance value, calculated for 1000 realizations of the Bright
Type I data that included the statistical and systematic errors in the
burst positions.

\bigskip
Figure 1b -- Same as in Figure 1a, except now for the 204 bursts
that represent the combined All Type I \& II Burst sub-class.

\bigskip
Figure 2 -- Logarithm of significance value for data sets with once
applied errors ($\log Q_R$) vs. the mean logarithm of significance for
sets with twice applied errors  ($\log Q_{RR}$). 100 burst data sets
were generated where each burst had a $20\%$ chance of repeating once.
Bursts were generated until a total of 204 bursts was reached. 100
values of $\log Q_{R}$ vs. $\log Q_{RR}$ were generated for each data
set.  The solid diamond corresponds to the mean values for the
$10,000$ runs, and the solid line corresponds to the best fit straight
line.  The open circles correspond to the experimental values of $\log
Q_M$ vs. $\overline {\log Q_M}$ for the most significant burst sub-classes.

\bye